\documentclass[twocolumn,showpacs,preprintnumbers,amsmath,amssymb]{revtex4}
\usepackage{graphicx}
\usepackage{dcolumn}
\usepackage{bm}

\begin{document}



\title{Lineshape measurement of an extreme-weak amplitude-fluctuating light source by the photon-counting-based second-order correlation spectroscopy}
\author{Hyun-Gue Hong, Wontaek Seo, Moonjoo Lee, Wonshik Choi, Jai-Hyung Lee, and Kyungwon An}
\affiliation{School of Physics, Seoul National University, Seoul, 151-747, Korea}
\email{kwan@phya.snu.ac.kr}
\date{\today}

\begin{abstract}
We demonstrate lineshape measurement of an extreme-weak amplitude fluctuating light source by using the photon-counting-based second-order correlation spectroscopy combined with the heterodyne technique. The amplitude fluctuation of a finite bandwidth introduces a low-lying spectral structure in the lineshape and thus its effect can be isolated from that of the phase fluctuation. Our technique provides extreme sensitivity suited for single-atom-level applications.
\end{abstract}

\maketitle

Lineshape is one of the most basic properties to be measured for
characterization of a light source. Examples are resonance
fluorescence lineshape of atoms and lineshape of lasers operating
below and above laser threshold \cite{Schawlow1958}, where the
lineshapes provide information on the light emission processes
\cite{Scully1997}.

Recent development of fundamental devices such as the micromaser
\cite{Meschede1985}, the microlaser \cite{An1994}, the
one-trapped-atom laser \cite{McKeever2003} and even single-photon
sources on demand \cite{Kuhn2002,McKeever2004} make lineshape
measurement of these rumdimentary light sources extremely compelling
to perform. However, such measurements call for ultra sensitive
techniques due to extremely low output intensity ($\sim$ subpicowatt) 
of these systems. Furthermore, such techniques should provide a way to extract intrinsic lineshape in the presence of 
technical noises such as cavity vibration and pump fluctuation.

The conventional techniques such as monochromator, Fabry-Perot-type optical
spectrum analyzer and electrical spectrum analyzer in heterodyne
configuration usually require rather descent
input powers to be used.
Among these conventional techniques, heterodyne spectroscopy (HS) provides the best sensitivity \cite{Hoffges}. One may hope to improve the sensitivity further by employing photon counting.
However, HS with an electronic spectrum analyzer operates in photo current mode (analog), not compatible with photon counting mode (digital).

Close examination of HS, however, reveals a way to
overcome this incompatibility. The heterodyne signal $I_h(t)$ is a
beat note of a source and a local oscillator (LO) field. What a spectrum analyzer does is to produce the power spectrum $\left|{\cal{I}}_h(\omega)\right|^2$ of an input signal $I_h(t)$. This is equivalent to calculating first an
intensity-intensity correlation function $\langle I_h(t)
I_h(t+\tau)\rangle$ with $\langle\ldots\rangle$ indicating a time
average and then taking a Fourier transform of it.
Furthermore, the intensity-intesity correlation function $\langle I_h(t) I_h(t+\tau)\rangle$ is nothing but the second-order correlation $g^{(2)}(\tau)$ of the heterodyne field.
The measurement technique of $g^{(2)}(\tau)$ is well established in quantum optics, employing photon counting technique, and thus measurement of $g^{(2)}(\tau)$ can be readily performed for extremely low-signal cases.

In this Letter, we demonstrate the validity and usefulness of the photon-counting-based second-order correlation spectroscopy (SOCS) in the measurement of the lineshape of an extremely weak source with both phase and amplitude fluctuations. In addition, the contribution from the phase fluctuation can be separated from that of the amplitude fluctuation simulating technical noises.



Let us write the time variation of the electric field, the spectrum of which we want to measure, as
$E(t)=E_{0}(t)e^{i[\omega_0 t+\phi(t)]}$,
where $\phi(t)$ is a randomly fluctuating phase of the field and
$E_{0}(t)$ is a slowly varying envelope.
In HS, the signal of interest, $E(t)$, is mixed with an LO field, given by
$E_L(t)=E_{LO}e^{i\omega_L t}$,
where $\omega_L$ is the carrier frequency of LO. It is assumed that the linewidth of LO is so small compared to that of the source to be measured that the LO can be regarded as a monochromatic field. The heterodyne signal or the beat note is then
\begin{eqnarray}
I_h(t)&=&{|E(t)+E_L(t)|}^2 \nonumber\\
&=&|E_0(t)|^2+|E_{LO}|^2\nonumber\\&&
+E_{L0}^* E_{0}(t)e^{i[\omega_B t + \phi(t)]}
+c.c.
\end{eqnarray}
where $\omega_B=\omega_0-\omega_{L}$.

The second-order correlation function of the beat note is given by \cite{jakeman1970}
\begin{eqnarray}
g^{(2)}_B(\tau)&=& \frac{\langle I_h^*(t)I_h(t+\tau)\rangle}
{\langle I_h^*(t)\rangle \langle I_h(t+\tau)\rangle} \nonumber\\&=&
1+2\frac{I_0
I_{LO}}{{(I_0+I_{LO})}^2}|g^{(1)}(\tau)|\cos\omega_B\tau\nonumber\\&&
+\frac{I_0^2}{{(I_0+I_{LO})}^2}\left[g^{(2)}(\tau)-1\right] \label{eq8}
\end{eqnarray}
where $I_0=\langle |E_0(t)|^2\rangle$, $I_{LO}=|E_{LO}|^2$,
$g^{(1)}(\tau)$ and $g^{(2)}(\tau)$
are the first and the second-order correlation functions of $E(t)$
itself. Note that $ g^{(2)}_B(\tau)$ oscillates at frequency
$\omega_B$ with its amplitude mainly determined by $ g^{(1)}(\tau)$
and modified near the origin by $g^{(2)}(\tau)$.

Amplitude fluctuation can be modeled by considering an infinite superposition of intensity modulations like
${|E_{0}(t)|}^{2}=I_{0}(1+ M \cos\Omega t) $, 
where the modulation depth $M$ is less than 1. The corresponding field amplitude can be written as
\begin{equation}
E_{0}(t)=\sqrt{I_0}(1+a_1e^{i\Omega t}+a_{-1}e^{-i\Omega t}+a_{2}e^{i2\Omega t}\cdots )
\label{eq:4}
\end{equation}
where $1\gg a_{1},a_{-1}\gg a_2 \cdots$ for $M \ll1$. This means that the spectrum will have symmetric sidebands around the carrier frequency. For a finite band for amplitude fluctuations we can write
\begin{equation}
{|E_{0}(t)|}^{2}=I_0 \left[1+ \int m(\Omega') \cos\Omega' t \; d\Omega' \right]
\label{eq:3p}
\end{equation}
 with a modulation-depth density $m(\Omega)$, and the resulting spectrum would contain a low-lying wing structure around the carrier.
 
\begin{figure}
\includegraphics[width=3.4in]{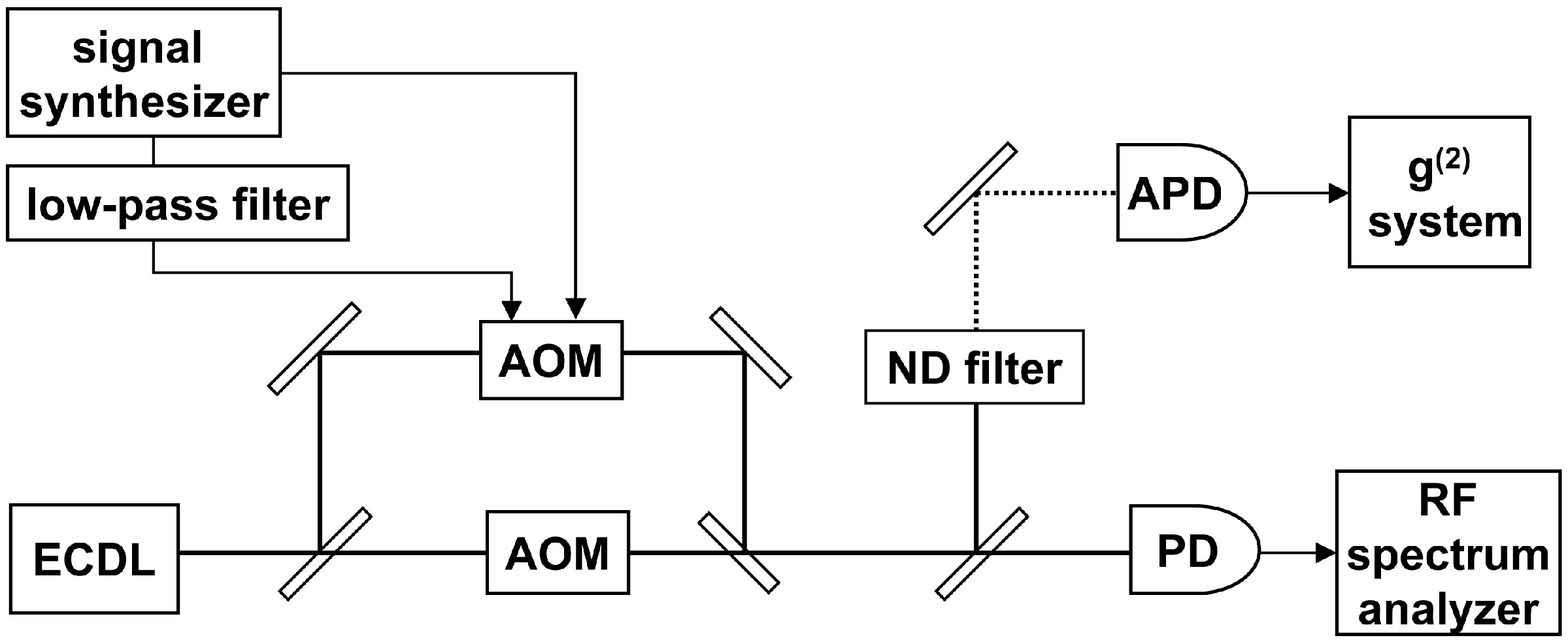}
\caption{Experimental scheme. The upper path corresponds to a test source and lower path corresponds to a local oscillator.}
\label{experiment}
\end{figure}

The schematic of our experiment is shown in Fig.\ \ref{experiment}.
A laser beam from an external cavity diode laser (ECDL) with a linewidth of about 0.5 MHz is split and guided into two acousto-optic modulators (AOMs), one of which is tunable (tuning rate of 22 MHz/V), imposing both phase and amplitude fluctuations to the laser beam. This beam is greatly attenuated and serves as a test light source whose spectrum is to be measured with a local oscillator, which is prepared by the other AOM with a fixed frequency shift.

Phase fluctuatuation is simulated by imposing a white noise of 1 MHz bandwidth from a signal synthesizer (Model DS345 by Stanford Research) on the frequency tuning port of the tunable AOM.
Amplitude fluctuation is provided by applying the white noise filtered by a low-pass filter (cutoff frequency of 50 kHz) to the intensity modulation port of the tunable AOM so that the intensity of the deflected beam off the AOM is modulated in a random manner.
The restricted bandwidth helps one to easily distinguish the effect of amplitude fluctuations from that of phase fluctuations in the experiment. For phase fluctuation no band filter is used since the white noise leads to a well-known Lorentzian lineshape.

Split beams are combined to generate a beat note. Since both the test light source and the local oscillator are derived from a common ECDL, the phase noise of the ECDL itself is exactly canceled out.
In addition the amplitude noise of the ECDL is negligible. Therefore, only the imposed phase and amplitude fluctuations determine the lineshape to be observed in the heterodyne field.

The combined beam is then split into two with a beam splitter. One beam leads to a silicon photodiode (Model FFD100 by Perkin Elmer) with
a minimum detectable power of 3.4 nW and the photo-current is analyzed
by an RF spectrum analyzer. The other is
attenuated by a set of neutral density filters down to the sub-picowatt
level, which is $ 10^{-9}$ of its original power and then detected
by an avalanche photodiode (APD, Model SPCM-AQR-13 by Perkin Elmer) in photon counting mode. The
mean photon counting rate is about 500 kcps. The
setup for measuring $g^{(2)}(\tau)$ is similar to that of our previous works 
\cite{Choi2005,Choi-PRL}.


\begin{figure}
\includegraphics[width=3.4in]{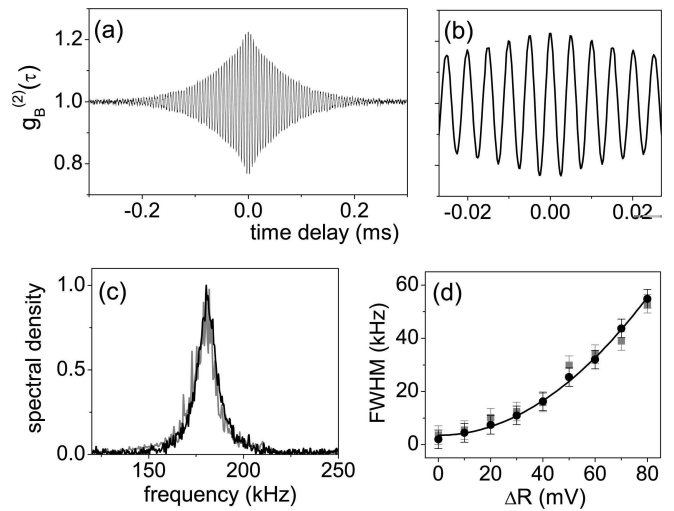}
\caption{(a) Observed $g^{(2)}_B(\tau)$ with phase fluctuations
only. Bin time: 250ns. (b) A detail view of the central region of (a). (c) Normalized spectral density in linear
scale. Gray line: conventional heterodyne spectroscopy, black line:
the second-order correlation spectroscopy. (d) Linewidth broadening
as a function of the noise rms voltage.} 
\label{fig3}
\end{figure}

Figure \ref{fig3} shows the result obtained when only the white
phase noise is present without amplitude noise. The decaying
envelope of $g^{(2)}(\tau)$ determines the linewidth while the
oscillation reflects the carrier frequency of the beat note. The
corresponding lineshape, obtained by taking the Fourier transform of
Fig.\ \ref{fig3}(a), is shown as a black curve in Fig.\
\ref{fig3}(c). A gray curve in Fig.\ \ref{fig3}(c) is obtained by
the conventional HS with the original power of
the test light source without the $10^{-9}$ attenuation. Note that
the spectra measured with two different methods are practically the same,
being a Lorentzian of a single linewidth. The full width at half
maximum (FWHM) is increased quadratically as the rms noise
voltage ($ \Delta R$) to the AOM is increased (Fig.\
\ref{fig3}(d)). The smallest linewidth, about 3 kHz, is determined
by the intrinsic noise in the AOM. Frequency resolution is
about 500 Hz for both lineshapes. The data acquisition time is 100
sec for both.

\begin{figure}
\includegraphics[width=3.4in]{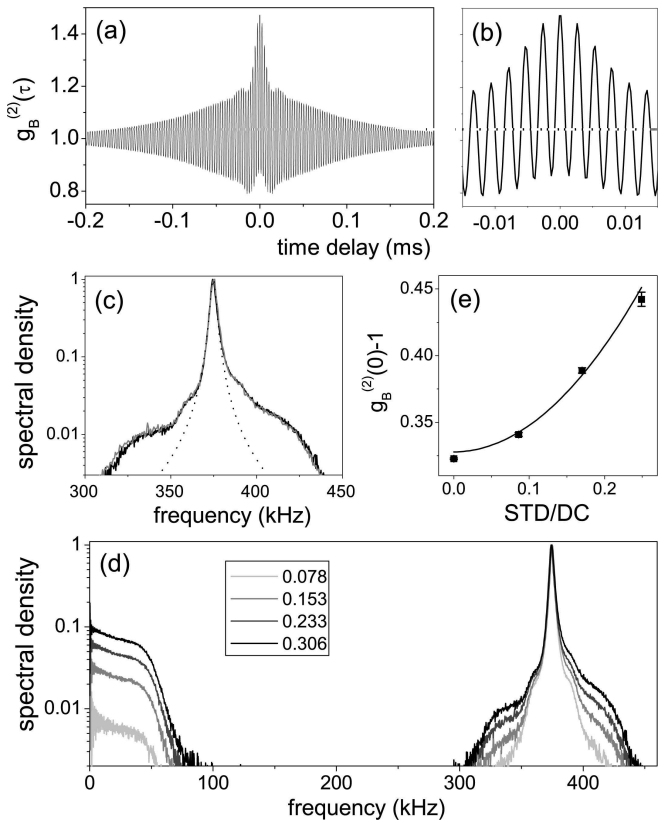}
\caption{(a) Observed $g^{(2)}_B(\tau)$ with amplitude fluctuations
included. (b) A detailed view of the central region of (a). (c) Normalized spectral density in log scale. The color
convention is the same as in Fig.\ \ref{fig3}(c). A Lorentzian fit (dotted line)
is shown for comparison. (d) Fourier transform of $g^{(2)}_B(\tau)$
as the ratio of rms to DC noise voltage is increased. (e) $g^{(2)}_B(0)-1$ as a
measure of amplitude fluctuation is plotted as a function of the
ratio of rms to DC noise voltage to the intensity modulation port of
the AOM.} 
\label{amspectrum}
\end{figure}

The effect of the amplitude noise is shown in Fig.\
\ref{amspectrum}. Intensity fluctuation causes an overall peak in $
g^{(2)}_B(\tau)$ near $ \tau=0$, corresponding to photon bunching
effect in quantum optics. The Fourier transform of $
g^{(2)}_B(\tau)$ is obviously not a Lorentzian of a single
linewidth.
It contains a low-lying wing structure reflecting the amplitude noise band function set by the electric low-pass filter.
The deviation from a Lorenztian (dotted line) is easily
recognizable in log-scale. Still the spectra measured with the two
different methods are practically identical, and thus the validity
of our $g^{(2)}$ spectroscopy for lineshape measurement for
extremely weak light source is well justified.

As the amplitude noise voltage is increased, FWHM does not show a
significant increase whereas the power spectral density of the low
lying structure does.
The ratio of rms to DC in the applied noise voltage to the AOM is
denoted by different depth of gray-scale in Fig.\
\ref{amspectrum}(d). The slight asymmetry observed in the spectral
density around the carrier frequency is due to the small leakage
voltage from the intensity modulation port to the frequency
modulation port of the AOM. 
The spectral structure
near the origin is also related to the spectrum of the imposed
amplitude noise (c.f. the third term in Eq.\ (\ref{eq8})).

From Eqs.\ (\ref{eq8}) and (\ref{eq:3p}), we can show that $g^{(2)}(0)-1$ proportional to the integral of the square of the modulation-depth density $m(\Omega)$ of the amplitude fluctuatuions \cite{Hong2006}. This relation is clearly seen in Fig.\ \ref{amspectrum}(e), where $g^{(2)}_B(0)-1$ is proportional to the square of the rms-to-DC ratio of the applied amplitude noise voltage.

It is noted that manifestation of the amplitude noise, usually of classical origin, as a low-lying spectral structure allows us to distinguish it from a Lorentzian lineshape due to the phase noise of often quantum origin like spontaneous emission in the laser. Although ultimate signal-to-noise ratios of both SOCS and the conventional HS are determined by the number of photons detected during the entire measurement time, the photon-counting based SOCS is observed to be $10^9$ times more sensitive than the current-based HS in our experiment since the former is free from thermal and other technical noises usually present in the latter. Our technique with an extreme sensitivity thus can be applied to measuring the lineshapes of the single-atom radiators mentioned above.

This work was supported by Korea Science and Engineering Foundation
Grants (NRL-2005-01371) and Korea Research Foundation Grants
(KRF-2005-070-C00058). 



\newpage

\end{document}